\documentclass{svproc}

\usepackage{url}
\usepackage{makeidx}         
\usepackage{graphicx}        
\usepackage{multicol}        
\usepackage[bottom]{footmisc}
\usepackage{booktabs} 
\usepackage{multirow}
\usepackage{tabularx}
\usepackage{algorithmic}
\usepackage{algorithm}
\usepackage{hyperref}
\usepackage{enumitem}
\usepackage{amsmath}
\usepackage{float}

\begin{document}
\mainmatter              
\title{\textit{Can Virtual Agents Care?} Designing an Empathetic and Personalized LLM-Driven Conversational Agent}
\titlerunning{Empathetic Virtual Agent Conversations}  
%
\author{Truong Le Minh Toan \inst{1, 2} \and Dieu Bang Mach\inst{3} \and Tan Duy Le\inst{4, 5, *} \and Nguyen Tan Viet Tuyen\inst{6}}
\authorrunning{T Truong et al.} 
%
%
\institute{
Swinburne University of Technology, Melbourne, Australia \and
Department of Computer Science, Swinburne Vietnam, FPT University, HCM city, 70000, Vietnam \href{FPT.EDU.VN}{FPT.EDU.VN} \and Deakin University, Melbourne, Australia
\and School of Computer Science and Engineering, International University, Ho Chi Minh City, Vietnam \and
Vietnam National University, Ho Chi Minh City, Vietnam \and  School of Electronics and Computer Science, University of Southampton, United Kingdom\\ *Corresponding author: ldtan@hcmiu.edu.vn}

\maketitle              

\begin{abstract}
Mental health challenges are rising globally, while traditional support services face limited availability and high costs. Large language models offer potential for conversational support, but often lack personalization, empathy, and factual grounding. A virtual agent framework is introduced to provide empathetic, personalized, and reliable wellbeing support through retrieval-augmented architecture, structured memory, and multimodal interaction. Objective benchmarks demonstrate improved retrieval and response quality, particularly for smaller models. A cross-cultural study with university students from Vietnam and Australia shows the system outperforms LLM-only baselines in coherence, perceived accuracy, and empathy, with most participants clearly preferring the proposed approach.
\end{abstract}

\section{Introduction}
Mental health and wellbeing challenges constitute a major public health crisis worldwide, affecting individuals across all demographics \cite{Magomedova2025MentalHealth}. The World Health Organization reports that nearly 1 in 8 people globally live with a mental health condition, with young people being disproportionately affected \cite{WHO2025OverBillionMentalHealth}. While demand for mental health support continues to rise, existing services face persistent structural, logistical, and resource constraints. These limitations have motivated growing interest in scalable, technology-driven solutions. Recent advances in large language models (LLMs) and retrieval-augmented generation (RAG) have enabled conversational agents with strong natural language capabilities and improved knowledge grounding \cite{WebGPTBrowserassistedQuestionanswering}. Although early commercial systems \cite{changAILedMentalHealth2024,WysaMentalWellbeing2020,prochaskaTherapeuticRelationalAgent2021} demonstrate potential for delivering cognitive behavioral therapy through dialogue, they remain limited by hallucinations, factual inaccuracies on sensitive topics, and shallow expressions of empathy. These challenges underscore the need for virtual agents that integrate reliable knowledge retrieval, personalized memory, empathetic interaction, and robust safety mechanisms within a unified framework.

In this paper, we present an empathetic and personalized virtual agent for wellbeing support that integrates a Context-Aware RAG pipeline, a dual-tier memory module, and a safety-filtering LLM. A Tri-Retrieval mechanism enhances information relevance, while multimodal virtual agent rendering improves engagement and trust. The primary contributions of this work are as follows:

\begin{enumerate}
    \item \textbf{Tri-Retrieval RAG Framework for Wellbeing Support}: A novel three-pronged retrieval strategy integrating keyword matching, semantic search, and real-time web retrieval to provide comprehensive, accurate, and up-to-date information grounding for wellbeing conversations.
    
    \item \textbf{Dual-Tier Memory System for Personalized Dialogue}: A memory retention mechanism that fuses short-term conversational context with long-term user interaction history through vector-based semantic retrieval.
    
    \item \textbf{Cross-Cultural Subjective Evaluation with Young Adults}: A comprehensive subjective evaluations is conducted with participants from Vietnam and Australia, validating the system's effectiveness in delivering empathetic and personalized wellbeing support across diverse cultural contexts.
\end{enumerate}

\section{Related Work}

Recent progress in large language models (LLMs) has significantly expanded the capabilities of open domain conversational agents. However, limitations in grounding, personalization, and empathetic responsiveness remain central challenges, particularly in systems intended for well being support or youth focused applications.

\subsection{Well-being chat bot} 
Early open-domain dialogue systems based on large transformers, such as BlenderBot \cite{rollerRecipesBuildingOpenDomain2021} and Meena \cite{freitasHumanlikeOpenDomainChatbot2020}, achieved substantial gains in fluency and coverage but often suffered from hallucination, factual drift, and inconsistencies without external knowledge. Empathetic and caring virtual agents, including Wysa \cite{changAILedMentalHealth2024,WysaMentalWellbeing2020} and Woebot \cite{prochaskaTherapeuticRelationalAgent2021}, illustrate how AI can provide psychologically informed guidance, yet they rely heavily on predefined scripts and do not adapt to long-term user preferences. However, most systems treat these separately, limiting personalization and failing to capture evolving habits, reasoning styles, or culturally grounded guidance.

\subsection{Retrieval-Augmented Generation (RAG)}
RAG has emerged as a core paradigm for improving factual reliability in open domain conversational agents by integrating external knowledge retrieval with large language models, markedly reducing hallucinations compared to purely parametric approaches. Early frameworks such as REALM \cite{guuREALMRetrievalAugmentedLanguage}and DPR\cite{karpukhinDensePassageRetrieval2020a} established the effectiveness of conditioning generation on retrieved evidence. Hybrid retrieval architectures combining dense and sparse signals further improve robustness under domain shift. Modern RAG systems extend this foundation with real-time tool use, adaptive retrieval, and self-reflective mechanisms, enabling chat bots to incorporate up-to-date or domain-specific information dynamically \cite{WebGPTBrowserassistedQuestionanswering}. However, challenges persist in evidence fusion, retrieval noise sensitivity, and resolving contradictions between sources, which are limitations that are especially problematic in wellbeing contexts. These gaps underscore the need for uncertainty-aware and context-sensitive retrieval strategies tailored for supportive dialogue.

\subsection{Virtual Agents in Human Computer Interaction}
Virtual agents are becoming central to multimodal human–computer interaction as researchers work to build trustworthy, engaging systems, with embodied conversational agents (ECAs) using facial expressions, gaze, gesture, and increasingly WebGL/WebXR rendering to create socially intuitive interactions that enhance trust and social presence \cite{FullArticleSystematic}. Building on early systems like REA and Max, recent work integrates speech prosody, emotion-driven animation, and real-time eye tracking so web-based characters can respond dynamically to user affect \cite{panEmotionalVoicePuppetry2023}, and studies show that synchronizing verbal and non-verbal channels increases satisfaction, lowers cognitive load, and boosts perceived credibility in supportive and educational contexts \cite{changImpactVirtualAgents2022}. Although advances in lightweight 3D rendering and browser-based motion retargeting have made such avatars scalable across devices, most systems still lack long-term user modeling and culturally adaptive reasoning, leaving them expressive but psychologically generic and limiting their effectiveness for well-being–centered interactions.

\section{Methodology}
\begin{figure}[!t]
    \centering
    \includegraphics[width=1\textwidth]{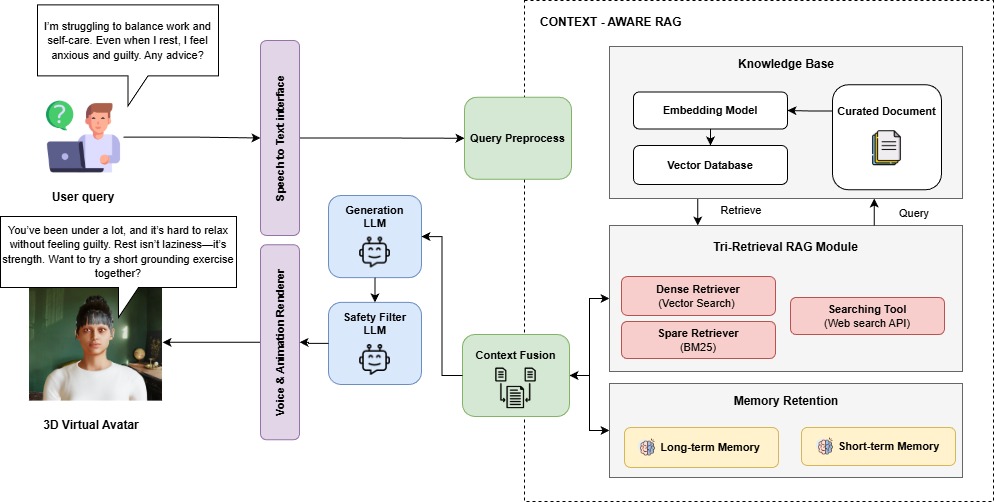}
    \caption{Overall framework of the Empathic and Personalized Conversation System. 
    The pipeline flows from user speech input to speech recognition, Tri-Retrieval Fusion (TRF), 
    empathetic LLM generation, memory integration, and virtual agent for expressive response.}
    \label{fig:system_architecture}
\end{figure}

The proposed system architecture is designed to facilitate high-fidelity, empathetic, and personalized conversational interactions through a virtual agent. The overall system operates on four key stages: user input processing, advanced content retrieval through Tri-Retrieval RAG module and memory retention, context fusion, and content moderation with multi-modal generation/rendering (as depicted in Figure \ref{fig:system_architecture}).

\subsection{Knowledge Base Construction}
The RAG system uses a multi-source knowledge base covering wellbeing topics from: (1) medical guidelines from Australian and Vietnamese agencies, (2) university wellbeing resources, and (3) peer-reviewed articles via Paperscraper \cite{paperscraper} (Table \ref{tab:dataset_overview}). Documents were chunked and embedded using Qwen3-Embedding-0.6B \cite{qwen3embedding}, which is he model ranks among the top performers on the Massive Multilingual Text Embedding Benchmark (MMTEB) \cite{embeddingRanking} while remaining highly efficient at 595M parameters. All embeddings were indexed in a vector database to support fast similarity search during retrieval

\begin{table}[!t]
\setlength{\tabcolsep}{6pt}  

\begin{tabularx}{\textwidth}{p{3.2cm}Xp{2.2cm}p{3cm}}
\hline\noalign{\smallskip}
Category & Sources & Format(s) & Size \\
\noalign{\smallskip}\hline\noalign{\smallskip}

Clinical \& Medical Guidelines 
& health.gov.au, healthdirect.gov.au, moh.gov.vn 
& PDF/HTML
& 45 docs, 780 chunks \\

Scientific Literature
& Academic databases (via Paperscraper)
& PDF
& 340 docs, 18{,}500 chunks \\

Institutional Wellbeing Resources
& University wellbeing centers; student counseling services
& HTML/PDF
& 52 docs, 920 chunks \\

\noalign{\smallskip}\hline\noalign{\smallskip}
\end{tabularx}
\caption{Document Categories and Dataset Composition}
\label{tab:dataset_overview}
\end{table}

\subsection{Context-Aware Retrieval and Multimodal Generation}

\subsubsection{Speech-to-Text Processing}

The speech-to-text module employs the Whisper model from Systran/faster-whisper-large-v3 \cite{whisper}, selected for its lightweight architecture that enables local device inference without requiring cloud-based processing, ensuring both privacy and real-time responsiveness.

\subsubsection{Tri-Retrieval Strategy}

To maximize information relevance and coverage while maintaining personalization and empathy, we introduced a three-pronged retrieval strategy named Tri-Retrieval (TR):

\textbf{Sparse Retriever (BM25):} Provides keyword-based retrieval to ensure high recall for specific or rare terms. BM25 represents an improvement over TF-IDF by addressing term frequency saturation and document length normalization issues, preventing bias toward longer documents \cite{BM25}.
    
\textbf{Dense Retriever (Vector Search):} Fetches information from the vector database based on semantic similarity, enabling sophisticated contextual understanding of user queries (e.g., recognizing emotional distress in statements such as “I feel overwhelmed but can't pinpoint why”). Similarity between query and document embeddings is computed using cosine similarity.

\textbf{Web Search API:} When Dense and Sparse retrieval fail to identify sufficiently relevant content, the system leverages real-time web search to ground responses in current information, preventing hallucination and ensuring factual accuracy for time-sensitive queries.

\subsubsection{Memory Retention for Personalization}

To enable personalized and contextually coherent interactions, we employ a dual-tier memory retention algorithm combining short-term dialogue context and long-term user history. Short-term memory stores recent turns to maintain session continuity, while long-term memory embeds all past interactions $(q_i, r_i)$ as vectors $e_i$ in a vector database $\mathcal{D}{long}$, capturing user preferences and recurring topics. For a new query $q_t$, its embedding $e_q$ is used to retrieve the top-$K$ most relevant historical interactions ($M{long}$), which are then fused with the short-term context $M_{short}$ to form the final memory state $M$. The full procedure is formalized in Algorithm~\ref{alg:memory}.

\begin{algorithm}
\caption{Memory Retention System}
\label{alg:memory}
\begin{algorithmic}[1]
\STATE \textbf{Input:} Current query $q_t$, conversation history $H = \{(q_1, r_1), ..., (q_{t-1}, r_{t-1})\}$
\STATE \textbf{Output:} Contextualized memory $M$

\STATE // Short-term Memory Management
\STATE $M_{short} \gets$ Extract last 5 turns from $H$
\STATE $M_{short} \gets \{(q_{t-5}, r_{t-5}), ..., (q_{t-1}, r_{t-1})\}$

\STATE // Long-term Memory Management
\FOR{each past conversation turn $(q_i, r_i) \in H$}
    \STATE $e_i \gets \text{Embed}(q_i \oplus r_i)$ \COMMENT{Concatenate and embed}
    \STATE Store $e_i$ in vector database $\mathcal{D}_{long}$
\ENDFOR

\STATE // Long-term Memory Retrieval
\STATE $e_q \gets \text{Embed}(q_t)$
\STATE $M_{long} \gets \text{TopK}(\mathcal{D}_{long}, e_q, k=3)$ \COMMENT{Retrieve top-3 similar conversations}

\STATE // Memory Fusion
\STATE $M \gets M_{short} \cup M_{long}$
\RETURN $M$
\end{algorithmic}
\end{algorithm}

\subsubsection{Response Generation with Safety Guardrails}

Retrieved knowledge from the Tri-Retrieval system and relevant memory context are fused with the user query and passed to the Generation LLM to produce empathetic, knowledge-grounded responses. The generated output is then evaluated by a Safety Filter LLM to prevent harmful, unethical, or inappropriate content. This filter operates using the following system prompt:

\begin{quote}
\textit{``You are a safety classifier. Evaluate the following response for: (1) encouragement or discussion of self-harm, suicide, or violence; (2) discriminatory, hateful, or toxic language; (3) inappropriate or unethical advice. Respond with 'SAFE' or 'UNSAFE' followed by a brief reason.''}
\end{quote}

\begin{figure}[!t]
\centering
\includegraphics[width=0.7\linewidth]{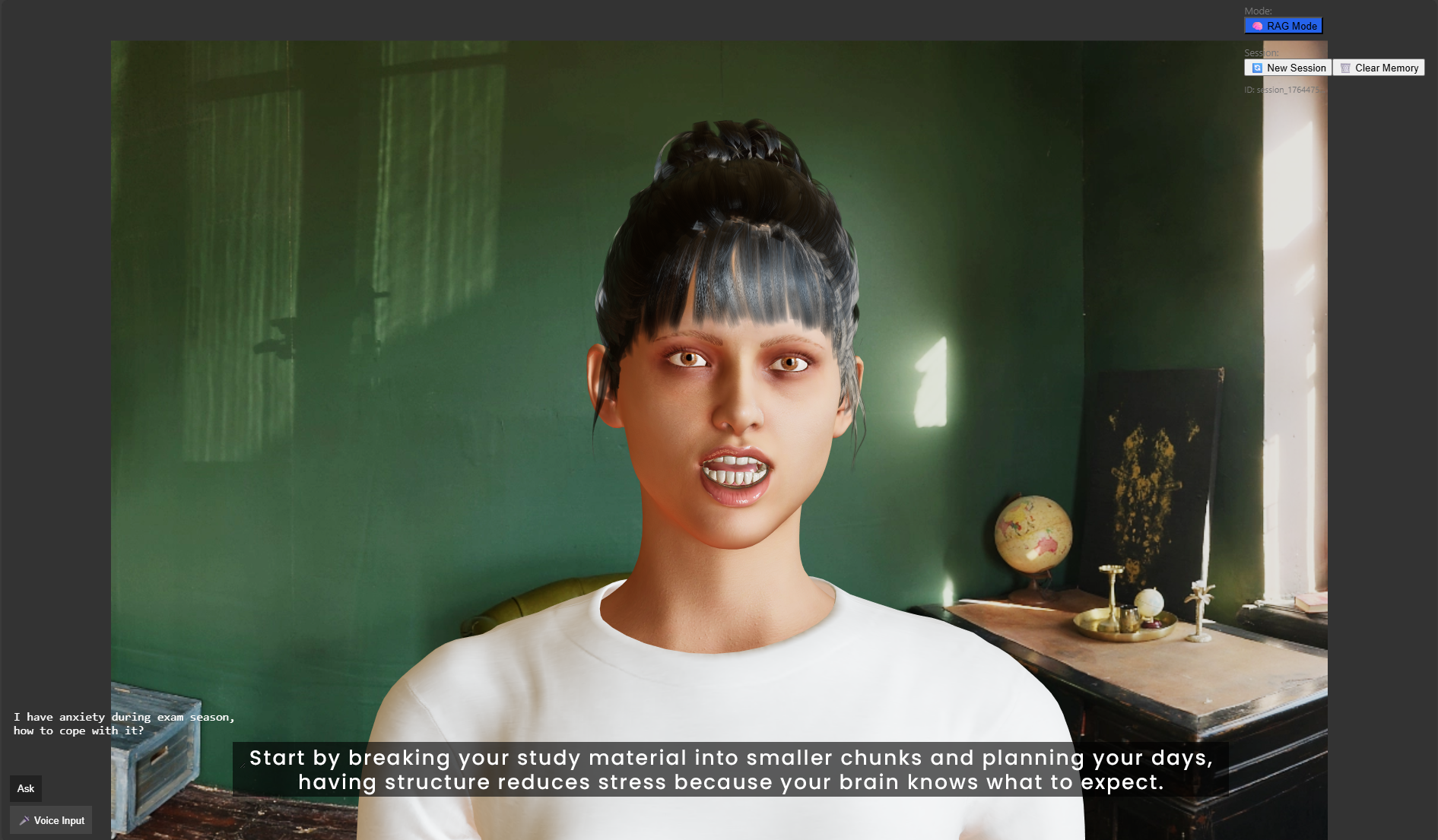}
\caption{Virtual Agent Interface with Real-Time Context-Aware Response Generation}
\label{fig:system_interface}
\end{figure}

Only responses classified as SAFE are delivered to the Voice \& Animation Renderer, which produces natural speech and synchronized facial expressions and gestures for the virtual agent. Figure \ref{fig:system_interface} illustrates the integrated user interface and overall system workflow.

\section{Experiment Evaluation}
\subsection{Retrieval Framework Performance}
\subsubsection{Context Acquisition Performance}

Retrieval effectiveness is measured with Precision@k and Recall@k, which assess the relevance of the top-k results and the proportion of relevant documents retrieved. Experiments use CLAP NQ \cite{clapnq}, a long-form QA benchmark with gold passage annotations that support precise alignment between retrieved documents and ground-truth evidence. Tri-Retrieval is evaluated against TF-IDF, BM25, and FAISS-based semantic retrieval, corresponding to its individual components. As shown in Table \ref{tab:retrieval_results}, it achieves the highest precision and recall at k=3 and k=5, outperforming all single-strategy baselines. At k=3, it reaches 0.635 P@3 and 0.742 R@3, improving on the best baseline by 5.3\% and 4.5\%. At k=5, it achieves 0.505 P@5 and 0.902 R@5, with gains of 3.9\% and 2.4\%. These results confirm that integrating lexical and semantic retrieval with web-augmented search captures complementary signals: BM25 supports exact matching, semantic retrieval captures conceptual similarity, and web search provides external knowledge coverage.

\begin{table}
    \centering
    \begin{tabular}{@{}lcccc@{}}\toprule
        \textbf{Method} & \textbf{P@3} & \textbf{R@3} & \textbf{P@5} & \textbf{R@5} \\\midrule
        Lexical Retriever (TF-IDF) & 0.443 & 0.529 & 0.370 & 0.750 \\
        Lexical Retriever (BM25)   & 0.543 & 0.629 & 0.470 & 0.850 \\
        Semantic Retriever (FAISS) & 0.603 & 0.710 & 0.486 & 0.881 \\
        \textbf{Tri-Retrieval}   \textbf{0.635} & \textbf{0.742} & \textbf{0.505} & \textbf{0.902} \\\bottomrule
    \end{tabular}
    \caption{Precision and recall results for different context acquisition approaches.}
    \label{tab:retrieval_results}
\end{table}

\subsubsection{Response Quality Evaluation}
Generation quality was assessed using F1, ROUGE-L, and BERTScore on SQuAD \cite{squad}, a widely used reading-comprehension benchmark with crowdsourced questions paired with Wikipedia passages, supporting rigorous assessment of both answer extraction and reasoning. Table \ref{tab:squad_results} shows Tri-Retrieval substantially improves all metrics, with GPT-4o achieving 0.7181 F1 (+79.5\% vs. zero-shot). Notably, LLaMA-3.2 improved 628\% (0.0666→0.4850 F1), demonstrating that retrieval grounding particularly benefits smaller models.

\begin{table}[h]
\centering
\begin{tabular}{llccccc}
\toprule
\textbf{Mode} & \textbf{Model} & \textbf{F1} & \textbf{ROUGE} & \textbf{BERTScore} \\
\midrule
\multirow{3}{*}{Tri-Retrieval}
& LLaMA-3.2 & 0.4850 & 0.4377 & 0.8919 \\
& GPT-3.5    & 0.5006 & 0.5139 & 0.9181 \\
& GPT-4o     & \textbf{0.7181} & \textbf{0.6964} & \textbf{0.9412} \\
\midrule
\multirow{3}{*}{Zero-shot}
& LLaMA-3.2 & 0.0666 & 0.0722 & 0.8173\\
& GPT-3.5    & 0.2111 & 0.2366 & 0.8633 \\
& GPT-4o     & \textbf{0.4000} & \textbf{0.5666}& \textbf{0.8706} \\
\bottomrule
\end{tabular}
\caption{Quality assessment of model responses across different answering modes.}
\label{tab:squad_results}
\end{table}
 
\subsection{Subjective Evaluation}
While objective metrics demonstrate retrieval accuracy and response quality, the ultimate success of a wellbeing support system depends on user experience and perceived trustworthiness. Prior work on conversational agents has established that users prioritize coherence, contextual relevance, and empathetic understanding when evaluating dialogue systems \cite{PintoBernal2025Designing}, with these factors being particularly critical in mental health and wellbeing contexts where user trust directly impacts therapeutic effectiveness. Recent studies on RAG-augmented chatbots\cite{robotSocial} suggest that grounding responses in external knowledge improves perceived accuracy, yet few have systematically evaluated whether these benefits translate to empathetic domains or generalize across diverse cultural contexts. Given that our target application serves young adults from different cultural backgrounds, it is essential to validate not only technical performance but also cross-cultural acceptability and user preference. To address these considerations, we conducted a comprehensive user study guided by the following research hypotheses:
\begin{enumerate}[label=\textbf{H\arabic*}, leftmargin=1.2cm]
\item \textbf{Effectiveness}: The RAG-augmented system provides significantly more helpful and contextually relevant responses for wellbeing queries than the LLM-only baseline.
\item \textbf{Coherence}: The RAG-augmented system maintains significantly better conversational coherence across multiple turns than the LLM-only baseline.
\item \textbf{User Perception}: Users perceive the RAG-augmented system as significantly more accurate, reliable, and understanding than the LLM-only baseline.
\item \textbf{Cross-Cultural Consistency}: The performance improvement of the RAG-augmented system over the LLM-only baseline remains consistent across Vietnamese and Australian participants.
\end{enumerate}

\subsubsection{Experimental Setup}
Twenty university students aged 20-30 were recruited from institutions in Vietnam (VNU-HCM, FPT University) and Australia (RMIT, Deakin, Swinburne), as shown in Table \ref{tab:participant_demographics}. This cross-cultural sampling strategy enables assessment of system effectiveness across diverse cultural contexts within the academic and research community.

\begin{table}[!t]  
\centering
\setlength{\tabcolsep}{7pt}  

\begin{tabularx}{\textwidth}{p{3cm}p{2cm}p{2.2cm}p{2cm}}
\toprule
\textbf{Demographic} & \textbf{Vietnam} & \textbf{Australia} & \textbf{Total} \\
\midrule
Total Participants & 11 & 10 & 21 \\
Age Range & 20--30 & 20--30 & 20--30 \\
Undergraduate & 7 & 5 & 12 \\
Postgraduate & 4 & 5 & 9 \\
Chatbot Experience & & & \\
\hspace{1em}-- None/Minimum & 3 & 2 & 5 \\
\hspace{1em}-- Moderate & 6 & 4 & 10 \\
\hspace{1em}-- Extensive & 2 & 4 & 6 \\
\bottomrule
\end{tabularx}
\caption{Participant Demographics}
\label{tab:participant_demographics}
\end{table}

Each participant used both system versions in a counterbalanced order to control for learning and fatigue. Version 1 (baseline) used an LLM-only setup without retrieval, while Version 2 (proposed) employed the full Tri-Retrieval RAG framework with memory and safety features. In each 10–15 minute session, participants held natural conversations spanning 3–5 wellbeing topics (e.g., stress, sleep, academic pressure, relationships, healthy habits. Immediately following each interaction, participants completed a structured questionnaire evaluating system performance across three dimensions including Effectiveness, Coherence, and User Perception using five-point Likert scales (1=Strongly Disagree to 5=Strongly Agree). After completing both sessions, participants indicated their preferred system version and provided open-ended qualitative feedback on their experience.

\begin{figure}[!t]
\centering
\includegraphics[width=0.7\linewidth]{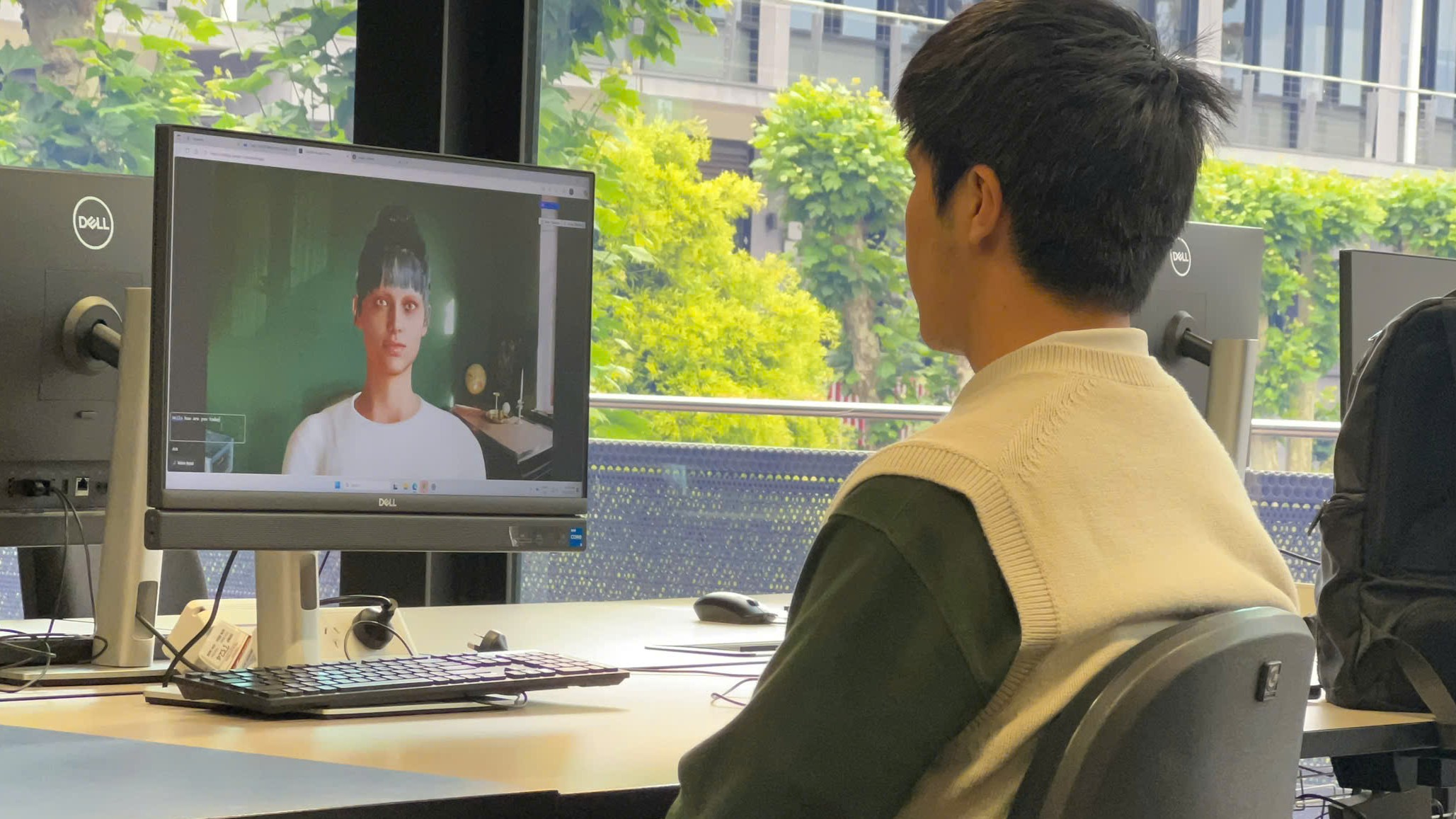} 
\caption{A participant interacting with the prototype during the evaluation session.}
\label{fig:user_testing}
\end{figure}

\subsubsection{Results and Analysis}
For each of the three aspects (Effectiveness, Coherence, and User Perception), composite scores were calculated as the mean of two corresponding items. Given the ordinal nature of the data and the within-subjects design, directional hypotheses were tested using one-tailed Wilcoxon signed-rank tests, with rank-biserial correlation ($r$) reported as the effect size.

\begin{table}[t]
\centering
\begin{tabular}{lccccccc}
\toprule
\textbf{Construct} & \textbf{RAG $M$ (SD)} & \textbf{LLM-only $M$ (SD)} & \textbf{$V$} & \textbf{$p$} (one-tailed) & \textbf{$r$} & \textbf{Sig.} \\
\midrule
Effectiveness      & 3.79 (0.64) & 3.57 (0.81) & 48.0  & .089 & .14 & n.s.       \\
Coherence          & 4.02 (0.60) & 3.02 (1.26) & 87.0  & .0018 & .52 & $^{**}$    \\
User Perception    & 3.86 (0.74) & 2.93 (0.87) & 129.0 & .0007 & .57 & $^{***}$   \\
\bottomrule
\end{tabular}
\caption{Subjective ratings of the RAG-augmented system versus the LLM-only baseline ($N=21$). Higher scores indicate more positive evaluation.}
\label{tab:subjective-results}
\begin{flushleft}
\small
Note. $^{**}$$p < .01$, $^{***}$$p < .001$ (one-tailed Wilcoxon signed-rank test).\\
Additionally, 90.5\% of participants (19/21) explicitly preferred the RAG-augmented version (binomial test against 50\%: $p < .001$).
\end{flushleft}
\end{table}

As shown in Table~\ref{tab:subjective-results}, the RAG-augmented system was rated significantly higher than the LLM-only baseline on Coherence ($p = .0018$, $r = .52$, medium-to-large effect) and User Perception ($p = .0007$, $r = .57$, large effect), fully supporting H2 and H3. The Coherence results demonstrate that the Tri-Retrieval strategy with memory retention successfully maintains contextual relevance across multiple conversation turns, while the strong User Perception scores indicate that participants found the system significantly more accurate, reliable, and empathetically understanding. Although the improvement in Effectiveness (H1) narrowly missed conventional significance ($p = .089$), the consistent directional trend combined with strong qualitative feedback suggests meaningful practical benefits. Open-ended responses reinforced these quantitative findings, with participants highlighting the RAG system's superior contextual awareness:

\begin{quote}
\small
``The RAG version was significantly better. It felt like a proper conversation partner because it could recall details like my degree and hobbies. The LLM-only mode was frustrating; it kept giving generic, surface-level advice without acknowledging the context I provided.'' (P14, Vietnamese)
\end{quote}

\begin{figure}[t]
\centering
\includegraphics[width=0.95\linewidth]{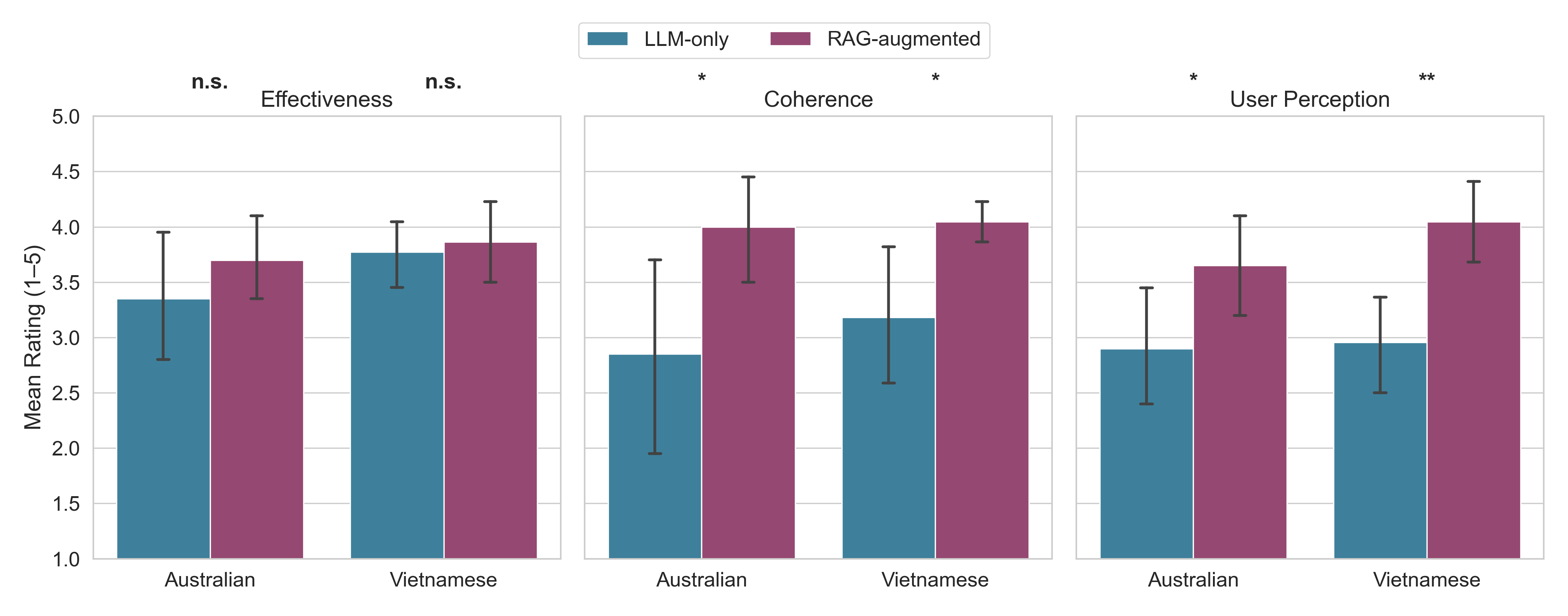}
\caption{Cross-cultural comparison of RAG improvement. Significant gains in Coherence and User Perception were observed in both Australian (n=10) and Vietnamese (n=11) participants, supporting H4.}
\label{fig:cross-cultural}
\end{figure}

Separate subgroup analyses confirmed H4 (cross-cultural consistency). As shown in Figure~\ref{fig:cross-cultural}, despite reduced statistical power ($n=10$ Australian; $n=11$ Vietnamese), the RAG system significantly outperformed the baseline on Coherence and User Perception in both cultural groups (Australian: $p = .031$ and $p = .027$; Vietnamese: $p = .023$ and $p = .0039$, respectively), yielding medium-to-large effect sizes ($r = .40$ - $.73$). Critically, these benefits manifested consistently across both cultures, with no meaningful differences observed in the pattern or magnitude of improvement. This cross-cultural robustness demonstrates that knowledge-grounded dialogue improves user experience independently of cultural background, which is a vital consideration for deploying mental health platforms. The non-significant Effectiveness result in both subgroups is attributable to ceiling effects and lower response variance rather than genuine cultural factors.

\section{Conclusion and Future Work}
This paper presents a novel virtual agent framework for empathetic and personalized wellbeing support, addressing critical limitations in existing conversational AI systems through the integration of a Tri-Retrieval RAG pipeline, dual-tier memory retention, and multimodal virtual agent rendering. Our empirical evaluation demonstrates that our proposed solution significantly improves both retrieval performance and response quality across multiple language models, with particularly strong gains for resource-constrained architectures. Cross-cultural subjective evaluation with 21 participants from Vietnam and Australia validates that the RAG-augmented system delivers significantly better conversational coherence and user perception of accuracy and empathy compared to the baselines, with these benefits generalizing consistently across both cultural contexts. The overwhelming user preference (90.5\%) for the RAG system, combined with qualitative feedback emphasizing superior contextual awareness and personalized responses, underscores the practical value of knowledge-grounded, memory-augmented dialogue for wellbeing applications.

Several limitations warrant attention: our participant sample remains limited to university students aged 20-30, restricting generalizability; the evaluation focused on short-term interactions (10-15 minutes), leaving long-term therapeutic efficacy unexplored; safety filtering has not been validated against adversarial inputs; and the system lacks comprehensive data encryption for stored conversations, raising privacy concerns for clinical deployment.

Future work should prioritize conducting longitudinal studies to assess sustained engagement and therapeutic outcomes, implementing robust data encryption and anonymization protocols, and validating safety mechanisms against adversarial inputs. Additionally, developing adaptive retrieval strategies that dynamically adjust based on conversation context and user emotional state could further improve system effectiveness. By addressing these limitations and building on the promising results demonstrated in this work, future iterations of empathetic virtual agents can provide scalable, accessible, and culturally sensitive mental health support.

%
%
\bibliographystyle{ieeetr} 
\bibliography{references}
\end{document}